\documentclass[final,5p,times,twocolumn]{elsarticle}

\usepackage{graphicx}

\usepackage{amssymb}

\journal{Nuclear Instruments and Methods A}

\begin{document}

\begin{frontmatter}



\title{Characterization and simulation of resistive-MPGDs with resistive strip and layer topologies}



\author[a]{J. Gal\'an}
\author[a]{D. Atti\'e}
\author[a]{A. Chaus}
\author[a]{P. Colas}
\author[a]{A. Delbart}
\author[a]{E. Ferrer-Ribas}
\author[a]{I. Giomataris}
\author[a]{F.J. Iguaz}
\author[a]{A. Gongadze}
\author[a]{T. Papaevangelou}
\author[a]{A.~Peyaud}
\address[a]{IRFU, CEA-Saclay, 91191 Gif-sur-Yvette, France}

\begin{abstract}
The use of resistive technologies to MPGD detectors is taking advantage for many new applications, including high rate and energetic particle flux scenarios. The recent use of these technologies in large area detectors makes necessary to understand and characterize the response of this type of detectors in order to optimize or constrain the parameters used in its production, material resistivity, strip width, or layer thickness. The values to be chosen will depend on the environmental conditions in which the detector will be placed, and the requirements in time resolution and gain, improving the detector performance for each given application. We present two different methods to calculate the propagation of charge diffusion through different resistive topologies; one is based on a FEM of solving the telegraph equation in our particular strip detector scheme, the other is based on a semi-analytical approach of charge diffusion and is used to determine the charge evolution in a resistive layer.
\end{abstract}


\end{frontmatter}



\section{Introduction}
One of the most prominent problems associated with gas-filled proportional chambers is the sparking induced by heavily ionizing particles producing large deposits. Amplified by the avalanche process they could reach a critical charge density, related to the Raether's limit \cite{bib1}, and could evolve into a discharge. As a result, a large fraction of the stored charge defining the amplification field is lost. The presence of discharges \emph{limits the high rate operation} of the detector due to the required power supply recovery time, \emph{reduces the mean life} of the detector and \emph{risks the damage of the readout electronics} due to the high currents reached by these spark processes.

To avoid any damage of the electronics, most of the detectors are using an additional protecting circuit, which interfaces the readout strips or pads with the front-end electronics. Recently new developments are undertaken to further improve the spark protection to operate at higher rates.

A promising approach is employing a resistive material on top of the anode plane \cite{fonte}. When a streamer develops the increasing number of charges is deposited at the high resistivity material. The anode potential generated by these charges reduces the amplification field, quenching the streamer resulting from field loss. A required condition is that the anode potential must last for a period of time long enough such that the charges present in the streamer have been completely evacuated, which introduces the need for high resistive values. The resistive interface limits the charge transfer, protecting the electronics, and avoids the total discharge of the mesh reducing the dead time of the detector.


\vspace{0.1cm}

This new technology needs to be characterized and simulated to exploit its possibilities, using the appropriate resistive and capacitive values depending on the needs of each particular application.

\vspace{0.1cm}

Micromegas detectors have evolved recently using this kind of protection. Two different topologies are typically implemented, a fully covering resistive layer and micromegas detectors with resistive strips. We present the description of these new approaches focusing on recent developments. Then, we will introduce a method to calculate the charge dispersion on the resistive material in each topology.

\subsection{Resistive strip topology}

\begin{figure}[htbp]
\begin{center}
\includegraphics[width=0.8\columnwidth,keepaspectratio]{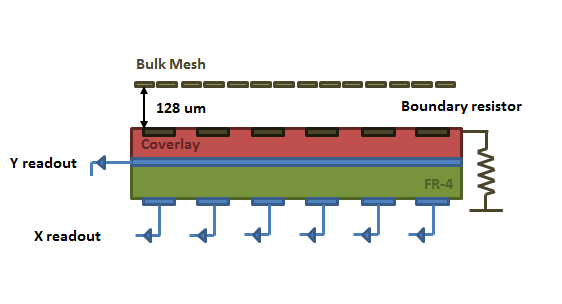}
\caption{Schema of the resistive micromegas detectors}
\label{fig1}
\end{center}
\end{figure}

For large Micromegas detectors, a novel protection scheme has been developed by a group working on hard radiation detectors at CERN \cite{bib7,bib8}. According to this scheme, we are using resistive strips above the readout electrode to overcome the spark problem. This technique consists in covering the micromegas readout~\cite{bib2,bib3} with a thin insulating layer where resistive strips are plotted on one favored axis (see Figure~\ref{fig1}). 

By adding a drift electrode, a MM detector operates as usually in the proportional avalanche mode inducing signals on the resistive strips plane. The resistive strip is grounded through an embedded high impedance resistor. The typical strip resistivity is in the order of 100 M$\Omega$/cm, and 200 M$\Omega$ for the boundary resistor which drives the strip to ground.

\subsection{Resistive layer topology}

\begin{figure}[htbp]
\begin{center}
\includegraphics[width=0.9\columnwidth,keepaspectratio]{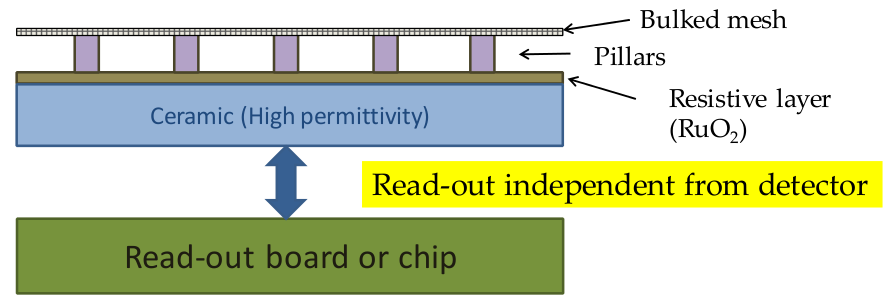}
\caption{Schema of the Piggyback detectors}
\label{fig2}
\end{center}
\end{figure}

An approach of resistive layer topology is called \emph{Piggyback} resistive Micromegas~\cite{piggy} where a thin resistive layer is deposited on an insulator. This element is used as substrate for the Micromegas which could be fabricated by using the \emph{bulk} technology~\cite{bulk}. A schematic view of the new structure is shown in figure~\ref{fig2} where we can see the various elements: the Micromegas structure with the pillars; the resistive layer; the insulating substrate and the readout board. The thickness of the resistive layer is of the order of 10 $\mu$m.

\vspace{0.1cm}
A particularity of this new detector technique is the fact that the readout electronics are independent (and thus exchangeable) from the micromegas structure. The elements of the structure are optimized in such a way that the electronic signal is not lost through the resistive layer but is propagated to a separate plane, carrying read-out pads or strips, by capacitive coupling. In order to optimize the induced signal by capacitive coupling the thickness of the insulating ceramic (t$_2$) must be kept small to satisfy the relation

\begin{equation}
\,\,\,\,\,\,\,\,\,\,\,\,\,\,\,\,\,\,\,\,\,\,\,\,\,\,\,\,\,\,\,\,\,\,\,\,\,\,\,\,\,\,\,\,\,\,\,\,  t_2 \ll t_1 \frac{\epsilon_2}{\epsilon_1},
\label{permRel}
\end{equation}

\noindent the amplification gap (t$_1$) is about 100 $\mu$m and it is achieved between a woven stainless steel mesh connected to the high voltage and a resistive anode. $\epsilon_1$ is the dielectric constant of medium 1 (gas) and $\epsilon_2$ is the dielectric constant of medium 2 (insulator). Because the insulator plays at the same time the role of vessel of the detector the thickness should be kept reasonable (several hundred $\mu$m). In order to satisfy (\ref{permRel}) the material should have a dielectric constant as high as possible. Good candidates are ceramic insulators having large dielectric constants ($\gg$10). The value for the amplification gap is the standard 128$\mu$m, while the ceramic thickness should be thicker than 300$\mu$m to assure the leak tightness of the detector. The resistive layer is made of ruthenium oxide (RuO$_2$). This material is extensively used for coating ceramics at high temperature for the preparation of resistors or integrated circuits, and provides good uniformity and a wide range of resistivities, between 1-100\,M$\Omega$/$\square$.



\section{Simulation of charge diffusion through a resistive strip}

A simplified approach of a single resistive strip is used in order to calculate the induced signal at the micromegas usual read-out and the charge diffusion through the resistive strip. The model here described is based on the equivalent electric circuit introduced by~\cite{resist}. This equivalent circuit describes in general a standard micromegas detector covered by a thin insulating layer where resistive strips are placed facing the standard metallic strips, creating an inter-capacitance and therefore an electric transmission line.

\vspace{0.2cm}

\begin{figure}[!ht]\begin{center}
\includegraphics[width=\columnwidth,keepaspectratio]{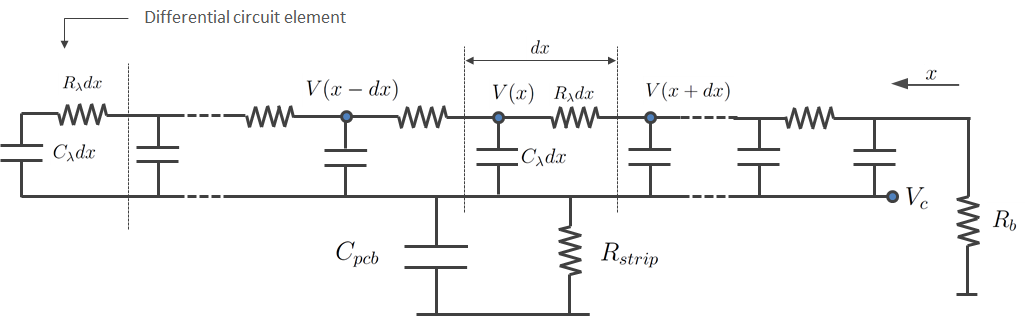}
\caption{Simplified RC-model of a resistive strip.}
\label{resistive_RCmodel}
\end{center}\end{figure}

In order to calculate the charge diffusion, the resistive strip is subdivided in a set of differential elements which final element is grounded through a boundary resistor $R_b$. The resistive strip has a linear resistivity $R_\lambda$ and is electrically coupled to the standard read-out, defined by $V_c$, through the mentioned insulating layer which introduces the strip linear capacitance $C_\lambda$. The standard read-out is connected in this model to a low impedance value $R_{strip}$, and the residual capacitance of the PCB board  $C_{pcb}$ (see figure~\ref{resistive_RCmodel}). 

The potential on the strip is driven by the following equation

\begin{equation}\label{eq:diffusion_1}
\,\,\,\,\,\,\,\,\,\,\,\,\,\,\,\,\frac{\partial^2 V(x,t)}{\partial x^2} = C_\lambda R_\lambda \frac{\partial\left( V(x,t) - V_c(t) \right) }{\partial t }.
\end{equation}

The evolution of the potential and/or charge in the resistive strip is bounded by the external circuit elements attached to the strip electrode, $R_{strip}$ and $C_{pcb}$, and the fixed resistive value, $R_b$, attached to the resistive strip. The first elements ($R_{strip}$ and $C_{pcb}$) allow determination of the strip voltage $V_c$ as a function of the resistive strip potential distribution, by integrating the total equivalent current induced at the capacitive surface $i_{C_\lambda}$ along the resistive strip length $x_L$,

\begin{equation}\label{eq1}
\begin{array}{cc}
\,\,\,\,\,\,\,\,\,\,\,\,\,\,i_{C_\lambda} = C_\lambda \int_0^{x_L} \frac{\partial}{\partial t} ( V(x,t)-V_c(t) ) dx = \\
\,\,\,\,\,\,\,\,\,\,\,\,\,\,= C_{pcb} \frac{dV_c(t)}{dt} + \frac{V_c(t)}{R_{strip}}.\,\,\,
\end{array}
\end{equation}

\vspace{0.2cm}

The resistive value $R_b$ provides the boundary condition on the voltage $V_o(t)$ at the beginning of the strip, and its value is related with the voltage gradient at the resistive strip end,

\begin{equation}
\,\,\,\,\,\,\,\,\,\,\,\,\,\,\,\,\,\,\,\,\,\,\,\,\,\,\,\,\,\,\,\,\,\,\,V_o(t) = -\frac{R_b}{R_\lambda} \frac{\partial V(x,t)}{\partial x} \Big |_{x=x_o}.
\end{equation}

\vspace{0.2cm}

The relation (\ref{eq:diffusion_1}) describes the potential propagation at the resistive strip. However, in order to observe such propagation we need a perturbation which is introduced as a time dependent current source spatially distributed and described by the change on linear charge density $\rho(x,t)$ along the resistive strip. This current is introduced inside relation (\ref{eq:diffusion_1}) by taking into account the expected voltage increase at the resistive nodes,

\begin{equation}\label{eq:diffusion_2}
\frac{\partial^2 V(x,t)}{\partial x^2} = C_\lambda R_\lambda \frac{\partial\left( V(x,t) - V_c(t) \right) }{\partial t } + R_\lambda \frac{\partial \rho (x,t)}{\partial t}
\end{equation}

\vspace{0.2cm}
\noindent the introduced charge density pretends to emulate the effect of charge induced by a physical avalanche, which would be the connection between our detector read-out and the physical signal.

\vspace{0.2cm}

The solution of the coupled relations (\ref{eq1}) and (\ref{eq:diffusion_2}) for a given input charge density $\rho(x,t)$ will provide a full description of the signal propagation within the frame of this model (a detailed solution can be found at~\cite{psd9}).

\begin{figure}[!ht]\begin{center}
\includegraphics[width=0.8\columnwidth,keepaspectratio]{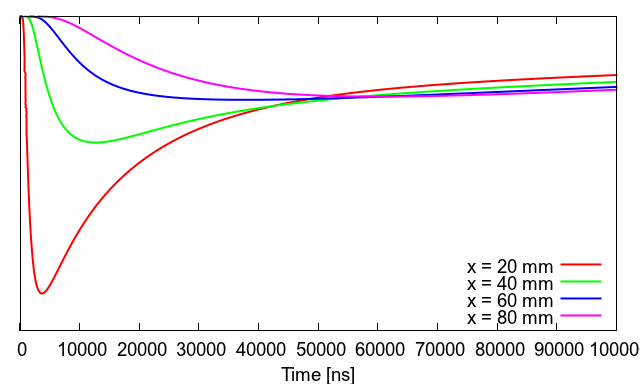}
\caption{Signal observed at the boundary resistor for currents applied at different resistive strip positions.}
\label{pulseShapes}
\end{center}\end{figure}

\vspace{0.1cm}
This model allows to solve particular scenarios and obtain the transition state of the strip receiving a charge stimulus. Figure~\ref{pulseShapes} shows the readout signal at the resistive boundary resistor and its shape dependency with a current applied at different strip positions for a total strip length of 10\,cm.

\begin{figure}[!ht]\begin{center}
\includegraphics[width=0.8\columnwidth,keepaspectratio]{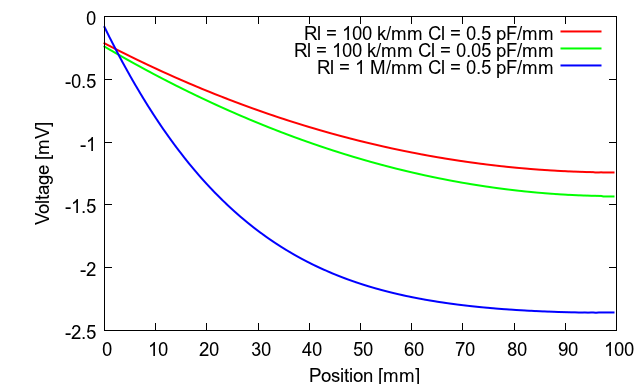}
\caption{Final potential distribution along the resistive strip for an homogeneous current distribution.}
\label{potDistr}
\end{center}\end{figure}

This method can be also used to obtain the potential profile in the resistive strip for different rate configurations. Figure~\ref{potDistr} shows the final state potential distribution resulting from an homogeneous rate distribution.

\section{Simulation of charge diffusion effects over a resistive plate}

In this section we show the effect on the gain due to the accumulation of charge over time due to slow charge diffusion in the illuminated region for a resistive layer topology. The charge diffusion relation for a given charge $Q$, which follows a Gaussian distribution of width $w$, was provided in~\cite{bib4}

$$
\rho(r,t) = \frac{Q}{2\pi ( 2ht + w^2 )} \mbox{exp}\left[ -\frac{r^2}{2 (2ht + w^2)} \right] \,\,\,\,\,\,\,\,\,\,\, h=1/RC
$$

\vspace{0.2cm}
\noindent where the parameter $h$ it is related to the surface resistivity $R$ and surface capacitance $C$. The accumulated charge will be translated into a surface potential given by the capacitance of the ceramic and thus, in a reduction of the amplification field. In order to study such reduction a narrow region around the Gaussian distributed charge has been chosen. The evolution on time of the total charge inside this region, defined by the radius $R_o$, it is integrated from the previous expression and is described by the following relation

$$
\rho(R_o,t) = Q \left[ 1 - \mbox{exp}\left( -\frac{R_o^2}{2 (2ht + w^2)} \right) \right].
$$

\vspace{0.2cm}
We can extrapolate this relation to the case of a continuous current flow by assuming this relation is valid for a differential element of charge. Integrating these differential contributions over time will allow us to calculate the accumulated charge at any time,

\begin{equation}
\begin{array}
{cc} Q(R_o,t_o) = \int_0^{t_o} \frac{d\rho(R_o,t)}{dt} dt = \\

= \int_0^{t_o} \left[ 1 - \mbox{exp}\left( -\frac{R_o^2}{2 (2h(t_o - t) + w^2)} \right) \right] \left( \frac{dq(V_a)}{dt} \right) dt
\end{array}
\end{equation}


\vspace{0.2cm}
\noindent where we should take into account that these differential charge contributions will be related to the local amplification field, modified by the anode potential $V_a$ at the resistive plane at each time interval.

\begin{figure}[htbp]
\begin{center}
\includegraphics[width=0.8\columnwidth,keepaspectratio]{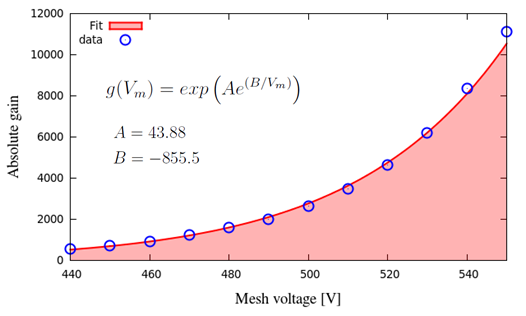}
\caption{Absolute gain measured (blue points) versus applied mesh voltage in Ar+10\% CO2, and the fitting result (red-filled curve).}
\label{fig10}
\end{center}
\end{figure}

\vspace{0.1cm}
The integral will be calculated by adding the charge contributions at each time interval n, and describing the gain in terms of the relative mesh versus anode potential,

$$
\delta q_n = g(V_a(n\delta t))N_e q_e r \,\,\,\,\,\,\,\,\,\,\,\,\,\,\,\,\, V_a(n\delta t) = q_n/\pi R_o^2 C
$$

\noindent where $N_e$ is the number of electrons produced in each interaction, $r$ is the interaction rate, $g$ is the gain as a function of the anode potential $V_a$. The gain as a function of the mesh potential is obtained experimentally by illuminating the Piggyback at low rates (see figure~\ref{fig10}), expression that will be used afterwards with the corrected potential given by V$_a$.

\vspace{0.1cm}

The charge is integrated considering the effect the accumulated charge will have in the gain at each time step. The simulation of each time step requires to know previously all the charges at each past time interval in order to apply the diffusion relation to them, this makes the computation of each time interval more expensive and the time required has a non-linear growth. The simulation of the first half an hour with reasonable accuracy can still be performed within a few hours.



\begin{figure}[htbp!]
\begin{center}
\includegraphics[width=0.8\columnwidth,keepaspectratio]{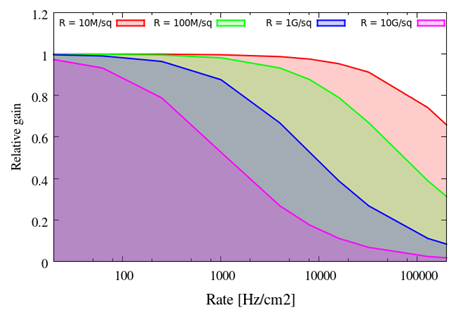}
\caption{ Relative gain versus the particle flux  for different resistivity values at 1pF/mm$^2$.}
\label{fig12}
\end{center}
\end{figure}

\vspace{0.1cm}
A set of simulations was launched at different scanning rates for different RC values in order to observe the effect on the gain drop. Figure~\ref{fig12} shows the final gain value obtained after an exposure of 5400 seconds, several simulations were performed at different rates in order to show the result as a function of the interaction flux. Different resistivity and capacitive values are tested from the range of 10M$\Omega/\square$  to 10\,G$\Omega/\square$, at 1\,pF/mm$^2$.

As lowest is the resistivity of the material, the longest is the flat gain region, and higher rates could be achieved without expecting an effect on the gain.

\begin{figure}[htbp]
\begin{center}
\includegraphics[width=0.9\columnwidth,keepaspectratio]{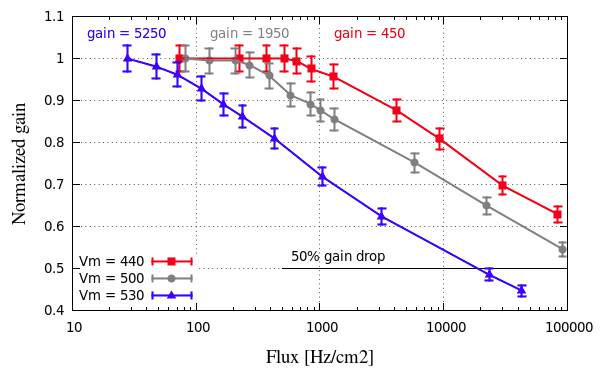}
\caption{Experimental gain drop dependency with the applied rate for different applied mesh voltages, $V_m$.}
\label{fig13}
\end{center}
\end{figure}

\vspace{0.1cm}
Experimental measurements performed with an X-ray generator show a behavior similar to the simulated one (see Figure~\ref{fig13}).

\section{Summary}

We report on two recent representative developments on resistive micromegas technology based on resistive layer and resistive strips topologies. These techniques profit from the resistive coating by increasing the mean life of the detector, reducing the sparks damage of the electronics and improving the rate capability by reducing the full discharge of the cathode.

We showed two methods which allows us to calculate the charge diffusion on the resistive material and predict the effects on the gain in each topology. The simulated effects are experimentally observed for the case of Piggyback detectors. The knowledge of the relation between the resistive values and the expected gain drop as a function of the rate will allow us to determine the optimum parameters to be used depending on the final application requirements.



\end{document}